\documentclass[twocolumn,showpacs,preprintnumbers,amsmath,amssymb]{revtex4}
\usepackage{dcolumn}
\usepackage{bm}
\usepackage{graphicx}
\usepackage{epsfig}

\begin{document}
\title{Taming Rogue waves in Vector BECs}

\author{P.S.Vinayagam$^1$}
\author{R.Radha$^1$}
\email{radha_ramaswamy@yahoo.com}
\author{K.Porsezian$^2$}
\email{ponzsol@yahoo.com}
\affiliation{$^1$ Centre for Nonlinear Science, PG and Research Dept. of Physics, Govt. College for Women (Autonomous), Kumbakonam 612001, India \\
$^2$Department of Physics, Pondicherry University,
Pondicherry-605014, India}

\begin{abstract}
Using Gauge transformation method, we generate rogue waves for the
two component Bose Einstein Condensates (BECs) governed by the
symmetric coupled Gross-Pitaevskii (GP) equations and study their
dynamics. We also suggest a mechanism to tame the rogue waves
either by manipulating the scattering length through Feshbach
resonance or the trapping frequency, a new phenomenon not
witnessed in the domain of BEC, we believe that these results  may
have wider ramifications in the management of rogons.
%
\end{abstract}

\maketitle
\section{Introduction}
\quad Rogue waves are nonlinear single oceanic waves of extremely
large amplitude, much higher than the average wave crests around
them and are localized both in space and time. Similar to the
properties of the solitary waves, rogue waves are also known as
'rogons' if they reappear virtually unaffected in size or shape
shortly after their interactions [1]. Rogue waves have caused
tremendous havoc and have contributed to several maritime
disasters[2]. In contrast to tsunamis [3] which can be predicted
hours (sometimes days) in advance, the danger of oceanic rogue
waves is that they appear from nowhere and disappear without even
a trace[4].

\quad Eventhough their existence has now been confirmed by several
observations, the grim reality is that their generating mechanism
is not yet fully understood. The recent studies argue that they
arise due to modulation instability [5,6,7], and their occurence
has been reported in optics [8], plasma [9] and Bose-Einstein
condensates (BECs)[10]. To date, several nonlinear partial
differential equations derived from different branches of physics
have been shown to admit rogue waves. Among these, the nonlinear
Schrodinger(NLS) equation represents the most elegant model to
describe rogue waves. Recently, using NLS equation, generating
mechanism for multi rogue waves has been proposed [11,12]. The
collision of two or more Akhmediev breathers (ABs) resulting from
the modulation instability can lead to rogue waves in these
systems [12]. At the same time, the discrete integrable systems
like generalized Ablowitz-Ladik-Hirota (ALH) lattice with variable
coefficients supports the nonautonomous discrete rogue solutions
[13]. Since the lifetime of rogue waves is very short, their
systematic investigation is very complicated. Penetrating deep
into the domain of rogue wave not only helps in understanding
their dynamics, but also in controlling their size and lifetime
for technological applications, particularly in the realm of
nonlinear optics and BECs.

\quad The recent theoretical investigations predict that the rogue
wave phenomenon can be observed in integrable multicomponent
systems like Manakov model [14], spinor F=1 condensates [15] etc.
and have also confirmed the existence of new type of bright-
dark-rogue wave solutions. The possible mechanism for the
formation of rogue waves in the two-dimensional coupled NLS
equations describing the nonlinearly interacting two-dimensional
waves in deep water has also been proposed [16-17]. The numerical
study on the two-component BECs with variable scattering lengths
shows that rogue wave solutions generated by phase and/or density
engineering can exist only for certain combinations of the
nonlinear coefficients describing two-body interactions [18].
Motivated by these observations, in this paper, we identify a
simple mechanism to generate and control the evolution of rogue
waves in vector BECs. It should be mentioned that even though
rogue waves have been manipulated in nonlinear optics [12,19] and
BECs [20], their observation in vector BECs characterized by the
symmetric coupled Gross-Pitaevskii (GP) equation has not yet been
fully understood. In this paper, we generate rogue waves for the
vector BECs governed by the coupled GP equation and manipulate
either the scattering length through Feshbach resonance or the
trapping frequency to tame them, a new phenomenon not observed in
the territory of BECs.

\section{Theoretical Model and Lax pair}

Considering a spinor BEC comprising of two hyperfine states $ |F=
1, m_{f}=-1>$ and $ |F= 1, m_{f}=1>$ of the same atom, say $^{87}
Rb $ [21] confined at different vertical positions by parabolic
traps, the dynamics in the mean field approximation is described
by the coupled dimensionless GP equation [22] of the following
form (for cigar shaped BECs)

\begin{eqnarray}
i \psi_{1t}+\psi_{1xx}+2\eta(t)(|\psi_1|^2
&+&|\psi_2|^2)\psi_1 + \lambda(t)^2 x^2 \psi_1 \nonumber\\
&+& i G(t)\psi_1 = 0,\\
i \psi_{2t}+\psi_{2xx}+2\eta(t)(|\psi_1|^2
&+&|\psi_2|^2)\psi_2 + \lambda(t)^2 x^2 \psi_2 \nonumber\\
&+& i G(t)\psi_2 = 0,
\end{eqnarray}

where,$\psi_j$(j=1,2) represents the order parameter of the
condensates, $\eta(t)$ is the temporal scattering length,
$\lambda(t)$ the trap frequency and $G(t)$ accounts for the
feeding of atoms (loss/gain) from the thermal cloud. The above
coupled GP equation has already been investigated [23,24] and the
collisional dynamics of bright solitons has been studied.

Eqs. (1),(2) admit the following Lax pair

\begin{eqnarray}
\Phi_x = Q_{1} \Phi\\
\Phi_t = Q_{2} \Phi
\end{eqnarray}
and
\begin{equation}
Q_{1}=
\left(%
\begin{array}{ccc}
  -i\zeta(t) & q_{1} & q_{2} \\
  -q_{1}^* & i \zeta(t) & 0 \\
  -q_{2}^* & 0 & i \zeta(t) \\
\end{array}%
\right),
\end{equation}
\begin{equation}
Q_{2} = \left(%
\begin{array}{ccc}
  Q_{2}(11) & Q_{2}(12) & Q_{2}(13) \\

  Q_{2}(21) & Q_{2}(22) & Q_{2}(23) \\

  Q_{2}(31) & Q_{2}(32) & Q_{2}(33)\\

\end{array}%
\right),
\end{equation}
where

\begin{eqnarray}
Q_{2}(11)&=& -2 i \zeta(t)^2 + 2 \zeta(t) i \Gamma(t) x+ i (|q_{1}|^2+|q_{2}|^2) \nonumber \\
Q_{2}(12)&=& 2 \zeta(t) q_{1}+ i (q_{1x}+2i \Gamma(t) x q_{1})\nonumber\\
Q_{2}(13)&=& 2 \zeta(t) q_{2}+ i(q_{2x}+2i\Gamma(t) x q_{2})\nonumber\\
Q_{2}(21)&=& -2 \zeta(t)q_{1}^*+ i (q_{1x}^*-2i\Gamma(t)x q_{1}^*)\nonumber\\
Q_{2}(22)&=& 2i\zeta(t)^2- 2\zeta(t)i\Gamma(t)x-i|q_{1}|^2 \nonumber\\
Q_{2}(23)&=& -i q_{2} q_{1}^* \nonumber\\
Q_{2}(31)&=& -2 \zeta(t)q_{2}^*+ i (q_{2x}^*-2i\Gamma(t)x q_{2}^*) \nonumber\\
Q_{2}(32)&=& -i q_{2}^* q_{1} \nonumber\\
Q_{2}(33)&=& 2 i \zeta(t)^2-2i\zeta(t)\Gamma(t) x-i |q_{2}|^2
\nonumber
\end{eqnarray}

\begin{eqnarray}
q_{1}(x,t)&=& \sqrt{\eta(t)} e^{(-i \Gamma(t) x^2/2)} \psi_1(x,t),\\
q_{2}(x,t)&=&\sqrt{\eta(t)} e^{(-i \Gamma(t) x^2/2)}\psi_2(x,t).
\end{eqnarray}
In eqs (3,4), $\Phi$ represents the eigenfunction denoted by
$(\phi_1, \phi_2, \phi_3)^T$ while $(Q_{1},Q_{2})$ denotes the Lax
operators described by $(3\times3)$ matrices while $\zeta(t)$
represents the nonisospectral parameter  obeying the following
equation
\begin{equation}
\zeta(t)=\mu  e^{(-2\int\Gamma(t)dt)}.
\end{equation}
In the above equation, $\mu$ is a complex constant and $\Gamma(t)$
is an arbitrary function of time. The compatibility condition
$Q_{1t}-Q_{2x}+ [Q_{1},Q_{2}]=0$ generates eqs.(1) and (2) with
the following constraints
\begin{equation}
G(t) = \Gamma(t) + \frac{1}{2}\frac{\eta'(t)}{\eta(t)},\nonumber\\
\end{equation}
and
\begin{equation}
\lambda(t)^2 = \Gamma(t)^2 +(\Gamma~'(t)/2).
\end{equation}
subject to the integrability condition
\begin{equation}
\lambda(t)^2 = G(t)^2 + \frac{1}{2}\frac{\eta'
(t)^2}{\eta(t)^2}-G(t)\frac{\eta'(t)}{\eta(t)}+\frac{1}{2}G'(t)-\frac{1}{4}\frac{\eta''(t)}{\eta(t)}.
\end{equation}

To obtain the exact solution of eqs (1, 2), we introduce the
following dependent variable transformation
\begin{equation}
\psi_{1}(x,t)=\Lambda(x,t) U(X,T),
\end{equation}
\begin{equation}
\psi_{2}(x,t)=\Lambda(x,t) V(X,T),
\end{equation}
with the coordinates governed by the following equations
\begin{eqnarray}
X&=& \sqrt{2} r_{0} \eta(t) x - 2\sqrt{2}b r_{0}^{3}\int
\eta(t)^{2}
dt \\
T&=&r_{0}^{2}\int \eta(t)^{2} dt
\end{eqnarray}
and
\begin{equation}
\Lambda(x,t)= \sqrt {2 r_{0}^{2} \eta(t)} e^{i
(-\frac{\eta(t)_{t}}{2\eta(t)} x^{2} + 2 b r_{0}^{2} \eta(t) x -2
b^{2}r_{0}^{4} \int \eta(t)^{2} dt)}
\end{equation}
where $r_{0}$ and $b$ are arbitrary constants so that eqs. (1,2)
reduce to the celebrated Manakov model.

\section{Construction of Rogue waves}

To construct rogue waves, we start from the following nonzero
plane wave solution as the seed solution given by

\begin{equation}
\psi_{1}[0]=c_{1} exp [i \theta_{1}],\psi_{2}[0]=c_{2} exp [i
\theta_{2}],
\end{equation}
where
\begin{eqnarray}
\theta_{1}=g_{1}x + (2 c_{1}^2+2 c_{2}^2 - g_{1}^2) t \\
\theta_{2}=g_{2}x + (2 c_{1}^2+2 c_{2}^2 - g_{2}^2) t
\end{eqnarray}

Feeding the above seed solution into the Lax-pair governed by eqs
(3,4), we obtain
\begin{eqnarray}
\Phi_{1x}=(M Q_{1} M^{-1}+M_{x}M^{-1})\Phi_{1}=\hat{Q_{1}}\Phi_{1}\nonumber\\
\Phi_{1t}=(M Q_{2} M^{-1}+M_{t}M^{-1})\Phi_{1}=\hat{Q_{2}}\Phi_{1}
\nonumber
\end{eqnarray}
where the iterated eigenfunction $\Phi_{1}=M \Phi, M=
diag[exp[-\frac{i}{3}(\theta_{1}+\theta_{2})],exp[\frac{i}{3}(2\theta_{1}-\theta_{2})],exp[\frac{i}{3}(\theta_{1}+\theta_{2})]]$
with
\begin{equation}
\hat{Q_1} =
\left(%
\begin{array}{ccc}
  \chi_{11} & c_{1} & c_{2} \\
  -c_{1} &  \chi_{22} & 0 \\
  -c_{2} & 0 &  \chi_{33} \\
\end{array}%
\right),
\end{equation}
\begin{equation}
\hat{Q_{2}}=i
\hat{Q_{1}}^{2}-[\frac{2}{3}(g_{1}+g_{2})-2\zeta_{1}]\hat{Q_{1}} +
m I
\end{equation}
\begin{equation}
\chi_{11}=-2i\zeta_{1}-\frac{i}{3}(g_{1}+g_{2})\nonumber
\end{equation}
\begin{equation}
\chi_{22}=i\zeta_{1}-\frac{i}{3}(2g_{1}-g_{2})\nonumber
\end{equation}
\begin{equation}
\chi_{33}=i\zeta_{1}+\frac{i}{3}(2g_{2}-g_{1})\nonumber
\end{equation}
and the new parameter\\ $m=2 i
\zeta_{1}^{2}+\frac{2i}{3}(c_{1}^{2}+c_{2}^{2}+\frac{2i}{9}(g_{1}^{2}-g_{1}
g_{2}+g_{2}^{2})+\frac{2i\zeta_{1}}{3}(g_{1}+g_{2})$.

In order to look for the rational solutions, we choose a new
parameter $\sigma=g_{2}+3 \zeta_{1R},
g_{1}=g_{2}-2\sigma,c_{1}=c_{2}=2\sigma$ where $g_{2}$ and
$\zeta_{1R}$ are arbitrary real numbers. The fundamental solution
matrix for Lax pair equations at $\zeta(t)=\zeta_{1}$ and
$\psi=\psi_j [0] (j=1,2)$ are $\Phi=M^{-1}\Theta$ where
\begin{equation}
\Theta=\left(%
\begin{array}{ccc}
 \phi_{11} & 4\sigma^{2}\nu +2\sqrt{3}\sigma & 4\sigma\\
 \phi_{21} & -2(\sqrt{3}-i)\nu-2\sigma & -2\sigma^{2}(\sqrt{3}-i)\\
 \phi_{31} & \phi_{22}^* & \phi_{23}^*\\
\end{array}%
\right)
\end{equation}
with
\begin{eqnarray}
\phi_{11}&=& 4\sigma^{2}(\nu + 2it)+4\sqrt{3}\sigma\nu+2 \nonumber \\
\phi_{21}&=& -2(\sqrt{3}-i)\sigma^2(\nu^{2}+2 i t)-4\sigma \nu \nonumber\\
\phi_{31}&=& -2(\sqrt{3}+i)\sigma^2(\nu^{2}+2 i t)-4\sigma \nu
\nonumber
\end{eqnarray}
and $ \nu= x + 2 \sqrt{3}(\sigma-i \sqrt{3} \zeta_{1R}) i t$. To
obtain the rational solution of the coupled GP equation, we
exploit the gauge transformation approach [25] employing the
following transformation

\begin{eqnarray}
\psi_{1}[1]=\psi_{1}[0] - 2i(\zeta_{1}-\bar\zeta_{1})\frac{\phi_1
\phi_2^*}{|\phi_1|^2+|\phi_2|^2+|\phi_3|^2}\nonumber\\
\psi_{2}[1]=\psi_{2}[0] - 2i(\zeta_{1}-\bar\zeta_{1})\frac{\phi_1
\phi_3^*}{|\phi_1|^2+|\phi_2|^2+|\phi_3|^2}\nonumber\\
\end{eqnarray}
The explicit forms of the first order rogue wave solution have the
following form

\begin{eqnarray}
\psi_{1}&=&\sqrt{\frac{2}{\eta(t)}}\varepsilon_1^{(1)}\beta(t)
\Big[-1-i\sqrt{3}+ a \Big]\times\nonumber\\
&& \exp [i \theta_{1}-\xi_{1}+\Gamma(t)x^{2}/2],\\
\psi_{2}&=&\sqrt{\frac{2}{\eta(t)}}\varepsilon_1^{(2)}\beta(t)
\Big[-1-i\sqrt{3}+ a \Big]\times\nonumber\\
&& \exp[i\theta_{2}-\xi_{1}+\Gamma(t)x^{2}/2]
\end{eqnarray}
where $a=f_{1}/f_{2}$
\begin{eqnarray}
f_{1}&=& -6\delta \sigma \sqrt{3}-36 t \sigma^{2}\sqrt
{3}-3+i(36 t \sigma^{2}+6\delta \sigma + 5\sqrt{3}),\nonumber\\
f_{2}&=& 12 \sigma^{2} \delta^{2}+8 \delta \sigma \sqrt 3 +144
t^{2} \sigma ^{4}+5\nonumber
\end{eqnarray}
where $\delta=x + 6 \zeta_{1R} t$. The gauge transformation
approach [25] can be easily extended to generate multi rogue wave
solution. For example, the second order rogue wave solution has
the following form
\begin{eqnarray}
\psi_{1}&=&\sqrt{\frac{2}{\eta(t)}}\varepsilon_1^{(1)}\beta(t)\Big[-1-i\sqrt{3}+
a1\Big]\times\nonumber\\
&&\exp [i \theta_{1}-\xi_{1}+\Gamma(t)x^{2}/2]\\
\psi_{2}&=&\sqrt{\frac{2}{\eta(t)}}\varepsilon_1^{(1)}\beta(t)\Big[-1-i\sqrt{3}+
a2\Big]\times\nonumber\\
&&\exp [i \theta_{1}-\xi_{1}+\Gamma(t)x^{2}/2]
\end{eqnarray}

where,
\begin{eqnarray}
a_1 &=& \frac{J_{1}+i K_{1}}{D}\nonumber\\
a_2 &=& \frac{J_{2}+i K_{2}}{D}\nonumber
\end{eqnarray}
\begin{eqnarray}
J_{1}=&-&864\sqrt{3}\sigma^{6}t^{3}-144\sqrt{3}\sigma^{5}\delta
t^{2}-72\sqrt{3}\sigma^{4}\delta^{2}t-216\sigma^{4}t^{2}\nonumber\\
&-&12\sqrt{3}\sigma^{3}\delta^{3}-144\sigma^{3}\delta t-18 \sigma^{2}\delta^{2}-12\sqrt{3}\sigma^{2}t+3\nonumber\\
J_{2}=&+&864\sqrt{3}\sigma^{6}t^{3}-144\sqrt{3}\sigma^{5}\delta
t^{2}+72\sqrt{3}\sigma^{4}\delta^{2}t-216\sigma^{4}
t^{2}\nonumber\\
&-&12\sqrt{3}\sigma^{3}\delta^{3}+144\sigma^{3}\delta t-18
\sigma^{2}\delta^{2}+12\sqrt{3}\sigma^{2}t+3\nonumber\\
K_{1}=&+&864\sigma^{6}t^{3}+144 \sigma^{5}\delta
t^{2}+72\sigma^{4}\delta^{2}t+312\sqrt{3}\sigma^{4}t^{2}+12\sigma^{3}\delta^{3} \nonumber\\
&+&96\sqrt{3}\sigma^{3}\delta t+18\sqrt{3}\sigma^{2}\delta^{2}+108
\sigma^{2}t+12\sigma\delta+\sqrt{3}\nonumber\\
K_{2}=&+&864\sigma^{6}t^{3}-144 \sigma^{5}\delta
t^{2}+72\sigma^{4}\delta^{2}t-312\sqrt{3}\sigma^{4}t^{2}-
12\sigma^{3}\delta^{3}\nonumber\\
&+&96\sqrt{3}\sigma^{3}\delta t-18\sqrt{3}\sigma^{2}\delta^{2}+108
\sigma^{2}t-12\sigma\delta-\sqrt{3}\nonumber\\
D=&+&1728\sigma^{8}t^{4}+384\sqrt{3}\sigma^{5}\delta t^{2}+12
\sigma^{4}\delta^{4}+432\sigma^{4}t^{2}\nonumber\\
&+&16\sqrt{3}\sigma^{3}\delta{3}+24\sigma^{2}\delta^{2}+4\sqrt{3}\sigma\delta+1\nonumber
\end{eqnarray}
where $\eta(t)= 2\sigma f(t)$, $\alpha(t)=\alpha_{0} \sigma
\exp[-2\int \Gamma(t) dt]$, $\beta(t)=\beta_{0} \sigma \exp[-2\int
\Gamma(t) dt]$.

\begin{figure}
\epsfig{file = 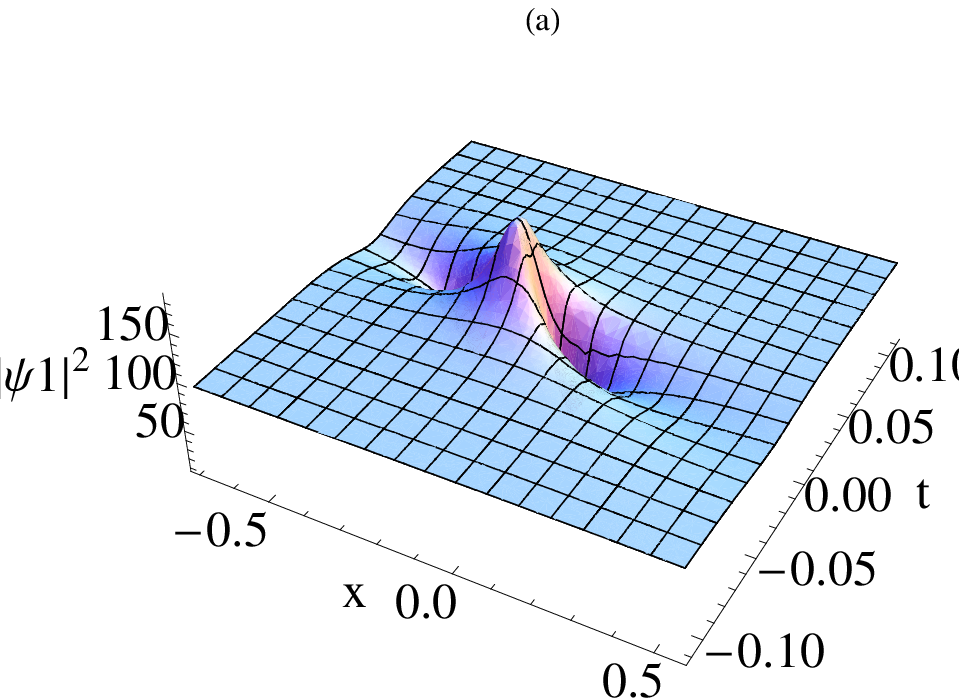,width=0.45\linewidth} \epsfig{file =
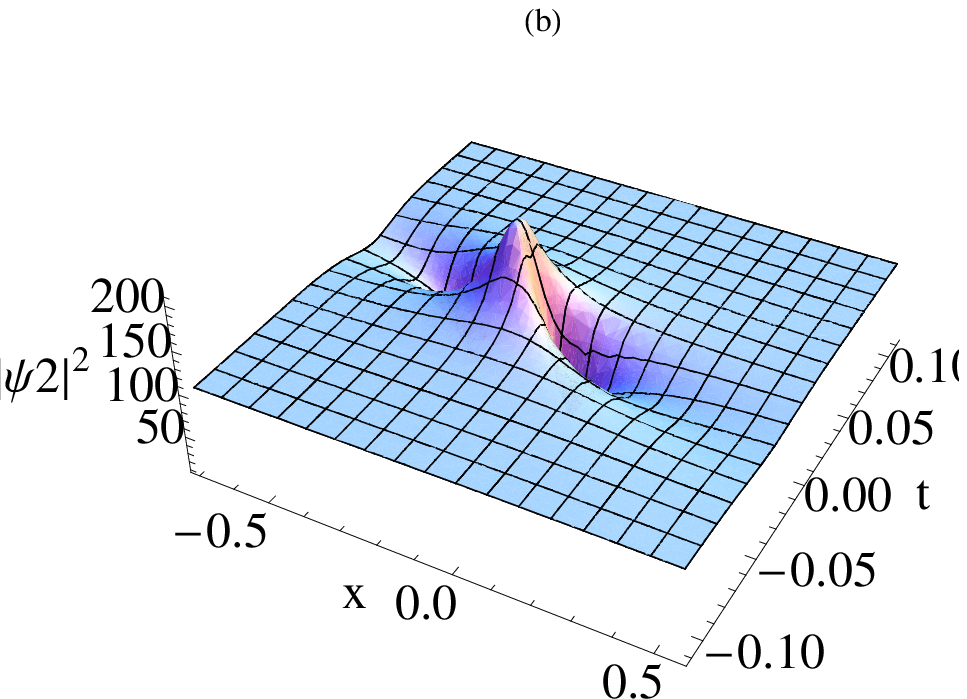, width=0.45\linewidth} \caption{Density profiles of
rogue waves for $\eta(t)=0.0006$, $g_{2}=0,
e_{1}=0.1,\Gamma(t)=0.01t,f(t)=0.001 $}
\end{figure}
\begin{figure}
\epsfig{file = 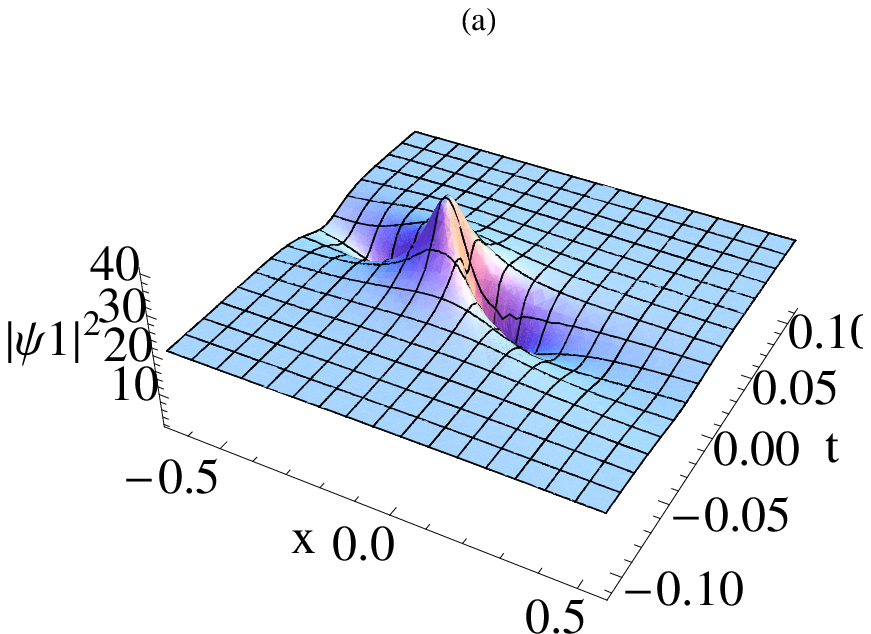,width=0.45\linewidth} \epsfig{file =
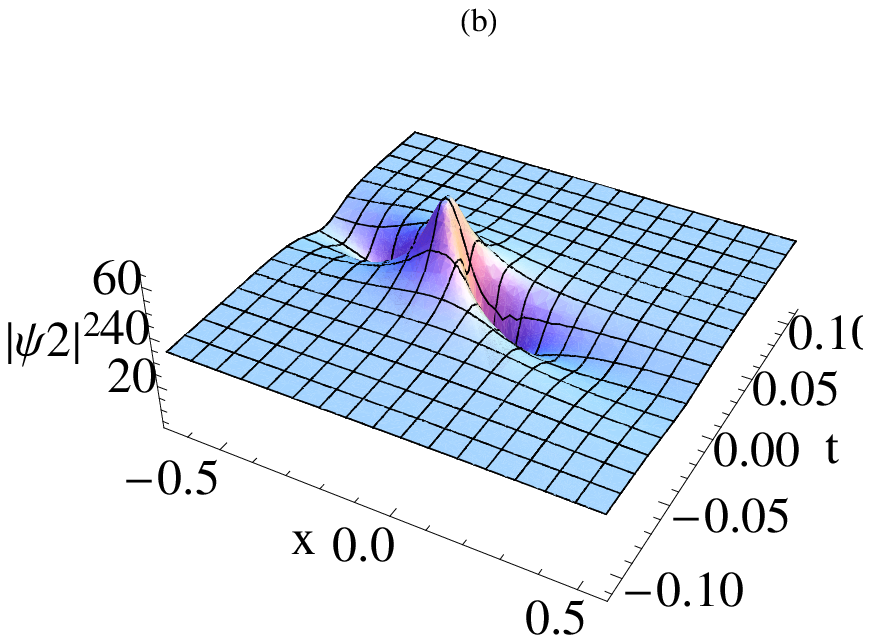, width=0.45\linewidth} \caption{Taming of the rogue
waves by manipulating the scattering length for $\eta(t)=0.006$
 and $f(t)=0.01$ with the other parameters as in fig 1. }
\end{figure}
\begin{figure}
\epsfig{file = 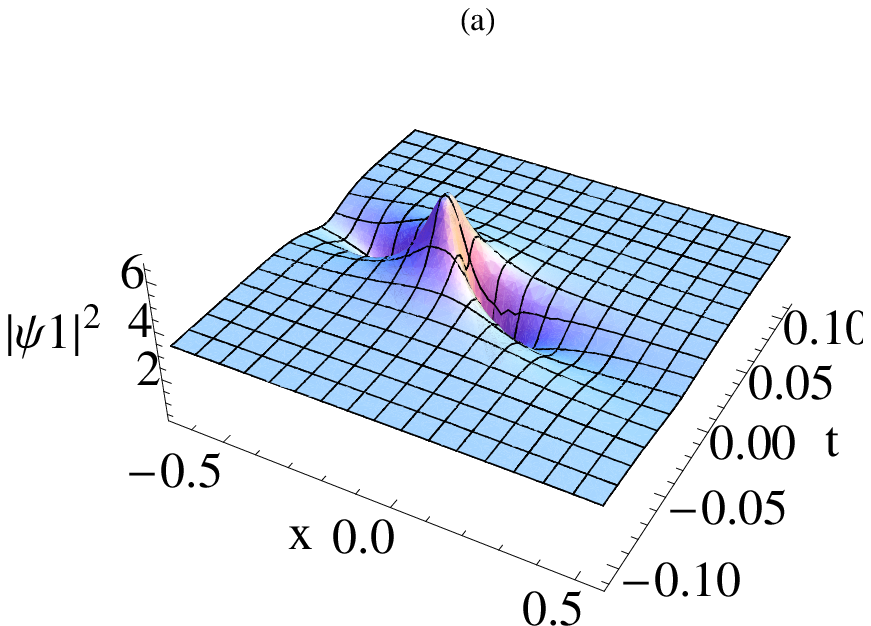,width=0.45\linewidth} \epsfig{file =
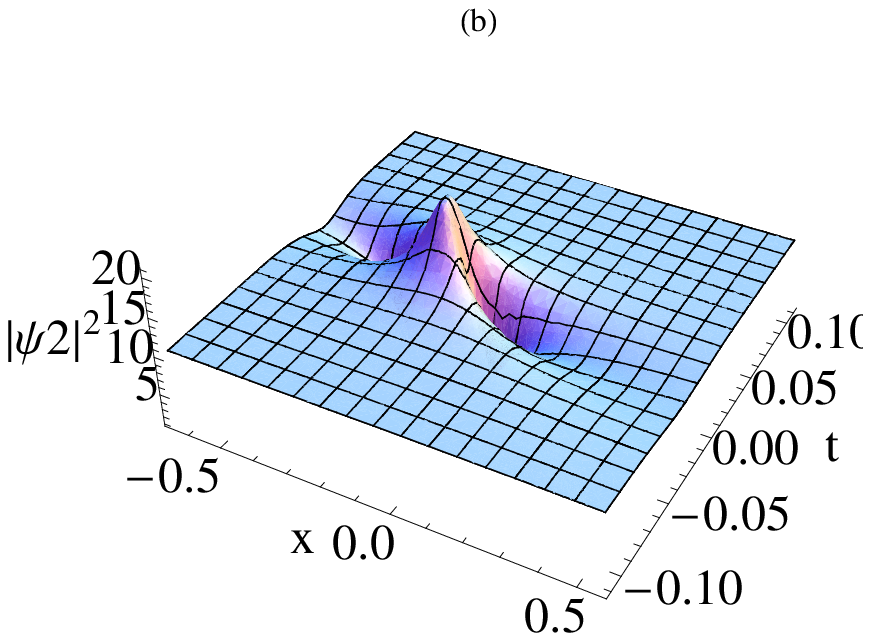, width=0.45\linewidth} \caption{Stabilization of rogue
waves for $\eta(t)=0.06$ and $f(t)=0.1$ with the other parameters
as in fig 1.}
\end{figure}
\begin{figure}
\epsfig{file = 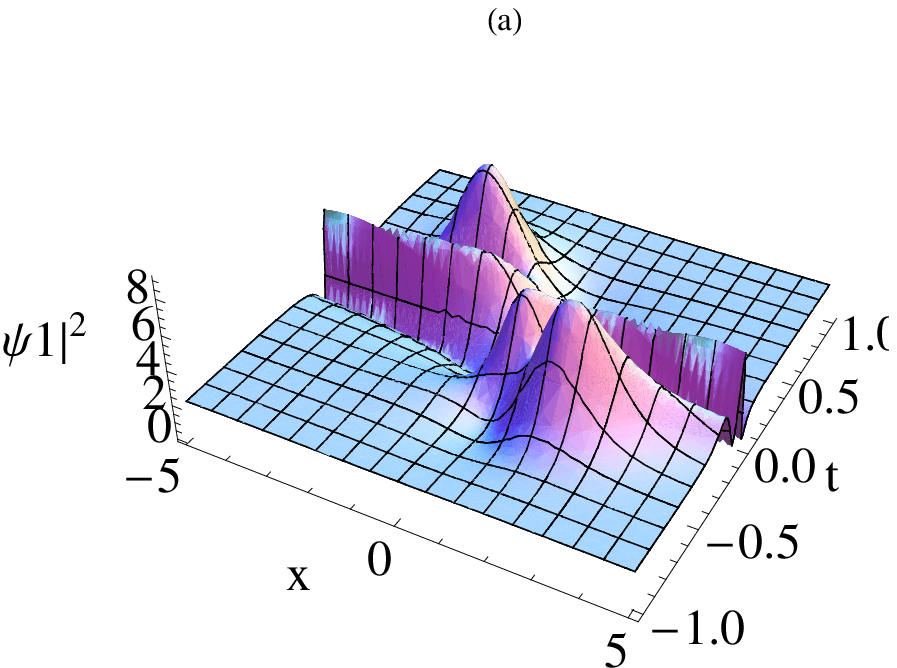,width=0.45\linewidth} \epsfig{file =
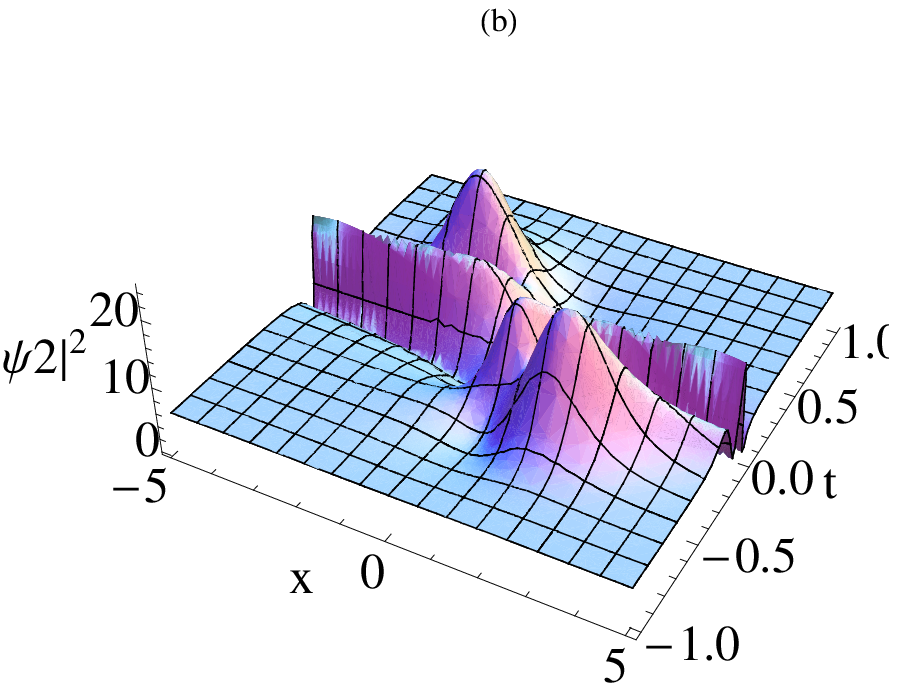, width=0.45\linewidth} \caption{Density profiles of two
rogue waves for $\eta(t)=0.12t $, $f(t)=0.05t$
$g_{2}=0.9,e_{1}=0.1,$ and $\Gamma(t)=0.1t,$ }
\end{figure}
\begin{figure}
\epsfig{file = 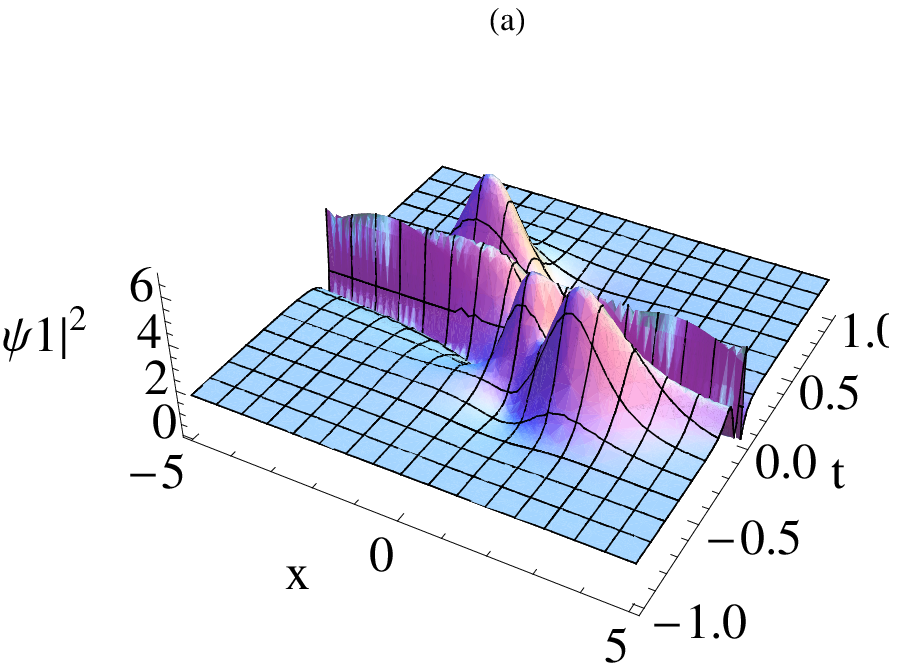,width=0.45\linewidth} \epsfig{file =
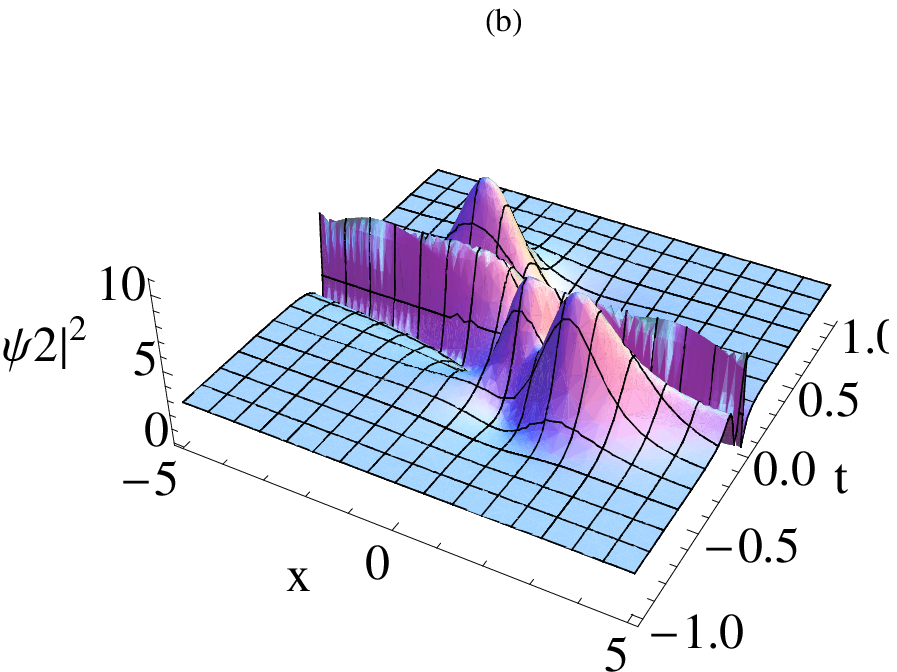, width=0.45\linewidth} \caption{Stabilization of two
rogue waves by manipulating the time dependent scattering length
for $\eta(t)=0.168t$ and $f(t)=0.07t$ with the other parameters as
in fig 4}
\end{figure}
\begin{figure}
\epsfig{file = 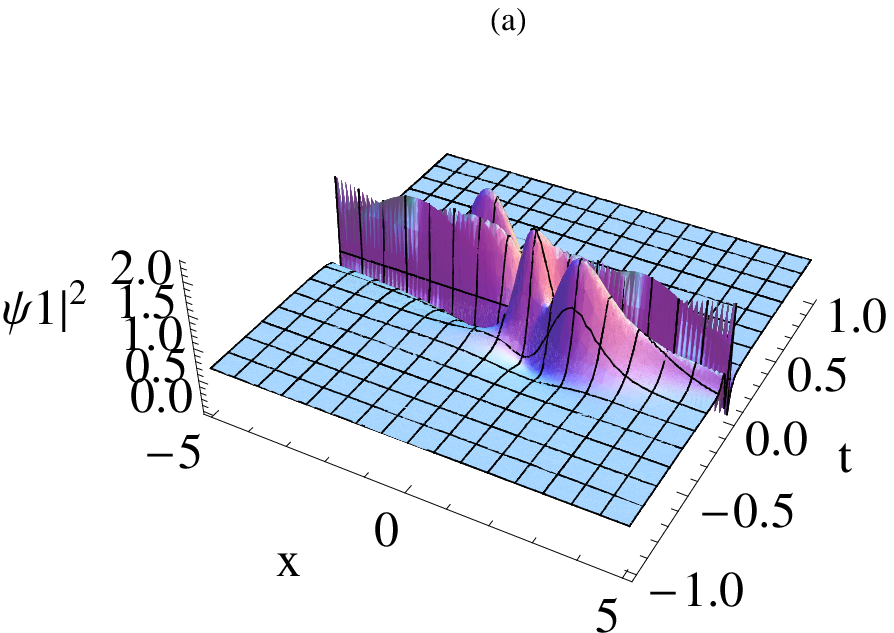,width=0.45\linewidth} \epsfig{file =
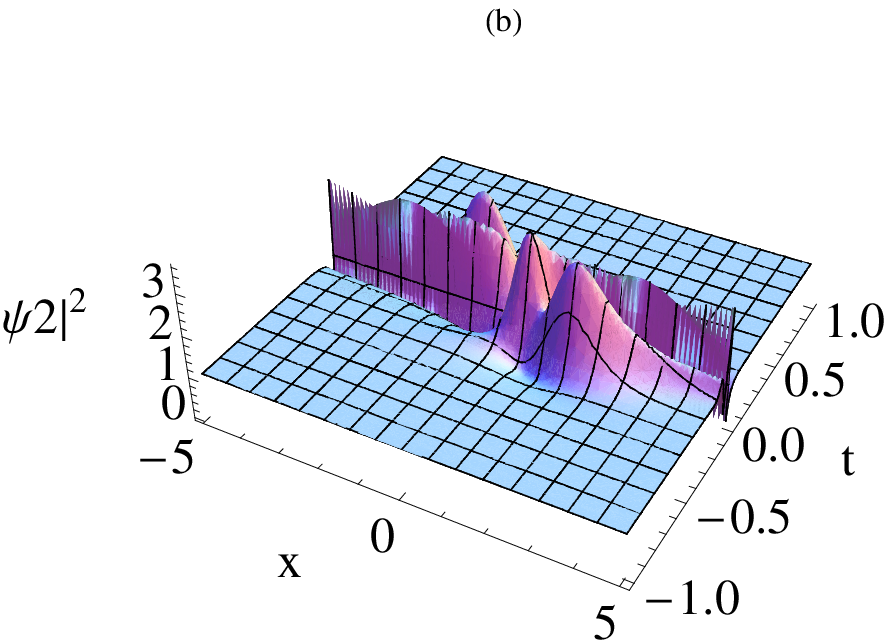, width=0.45\linewidth} \caption{Stabilization of two
rogue waves by finetuning the time dependent scattering length for
$\eta(t)=0.36t $ and $f(t)=0.15t$ with the other parameters as in
fig 4}
\end{figure}
\section{Stabilization of Rogue waves}
Fig.(1) shows the density profiles of first order rogue waves
governed by the scattering length for $\eta(t)=0.0006$ and
$f(t)=0.001$. It is obvious from fig (1) that the density of rogue
waves is enormous which means that it would collapse or disappear
in a short interval of time during time evolution. To stabilize
(reduce the density) the rogue waves and thereby increase its
lifespan, we harness the fact that their densities $|\psi_{j}|^2
(j=1,2)$ is inversely proportional to the scattering length
$\eta(t)$ (of course, $\eta(t)$ varies directly with $f(t)$).
Hence, we manipulate (increase) the scattering length $\eta(t)$
through Feshbach resonance suitably to stabilize the first order
rogue waves as shown in fig(2). Rogue waves can be stabilized
further for $\eta(t)=0.06$ and $f(t)=0.1$ as shown in fig(3). This
process of stabilizing the amplitude of rogue waves and thereby
increasing the lifetime is called 'Taming'. Fig (4) shows the
density profile of second order rogue waves for time dependent
scattering lengths $\eta(t)=0.12t$ and $f(t)=0.05t$. Again, one
can tame the rogue waves further by fine-tuning the time dependent
scattering lengths as shown in figs (5) and (6). From figs(4-6),
one understands that the density of the rogue waves decreases by
finetuning the time dependent scattering lengths. This means that
one can delay the inevitable (the collapse or disappearance of the
condensates) by manipulating the time dependent scattering lengths
as well. In addition, the fact that it stretches over a finite
interval of time compared to figs(1-3) means that one finally ends
up increasing the lifespan of BECs.

\quad It should also be mentioned that the trapping frequency
$\lambda(t)$ which is related to $\Gamma(t)$ by virtue of eqn.
(10) can also suitably changed to tame rogue waves. Fig (7) shows
the density profile of second order rogue waves for periodically
varying scattering lengths $\eta(t)=2 cos(0.15t)$ and
$f(t)=cos(0.15t)$ and fig.8 depicts the corresponding contour
plot. The contour plot shown in fig.9 depicts the time evolution
of second order rogue waves shown in fig.8 by maneuvering the trap
frequency $\Gamma(t)=0.1 t$. Further evolution of the second order
rogue waves shown in fig.10 shows that one can certainly enhance
the lifespan of the rogue waves by manipulating the trap frequency
for $\Gamma(t)=0.03t$

\begin{figure}
\epsfig{file = 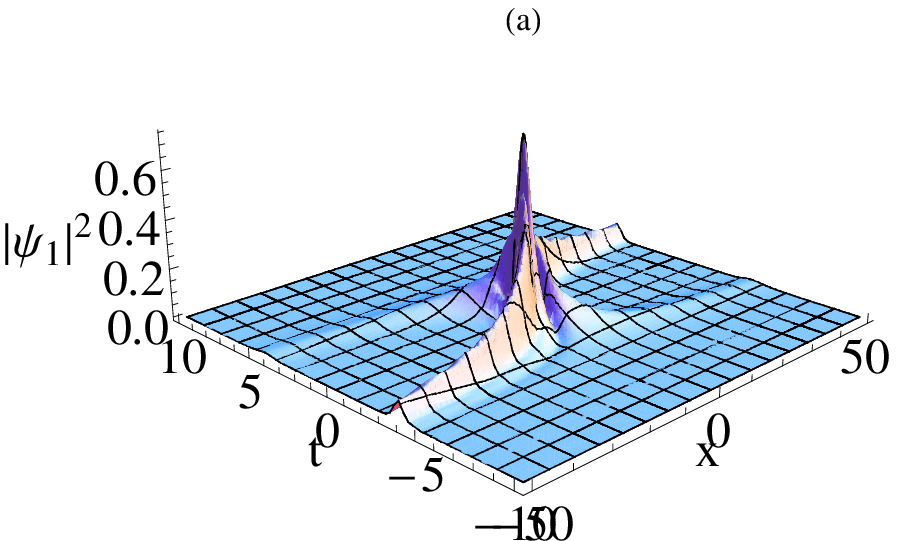,width=0.45\linewidth} \epsfig{file =
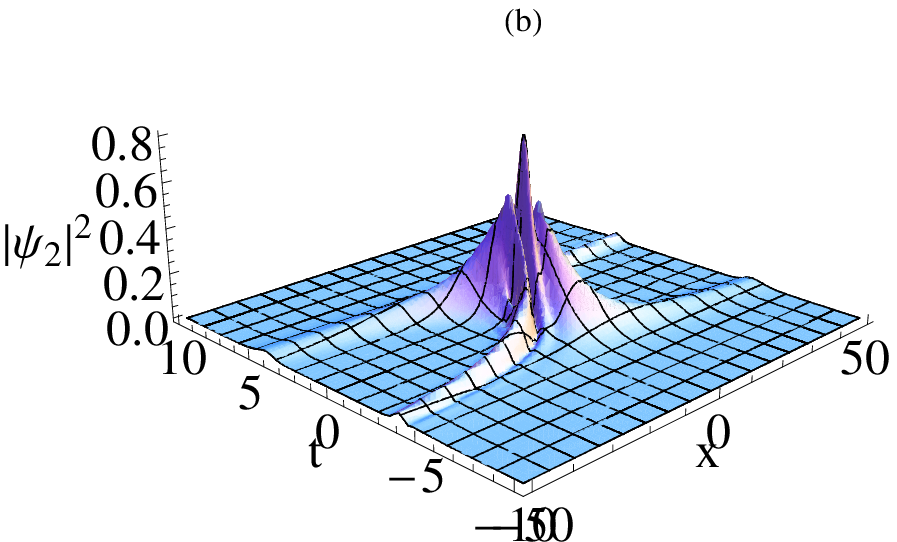, width=0.45\linewidth}\caption{Density profile of two
rogue waves for $\eta(t)=2cos(0.15t), g_{2}=0.5,
e_{1}=\frac{0.5}{3}, \Gamma(t)=0.15t,$ and $f(t)=cos(0.15t)$}
\end{figure}
\begin{figure}
\epsfig{file = 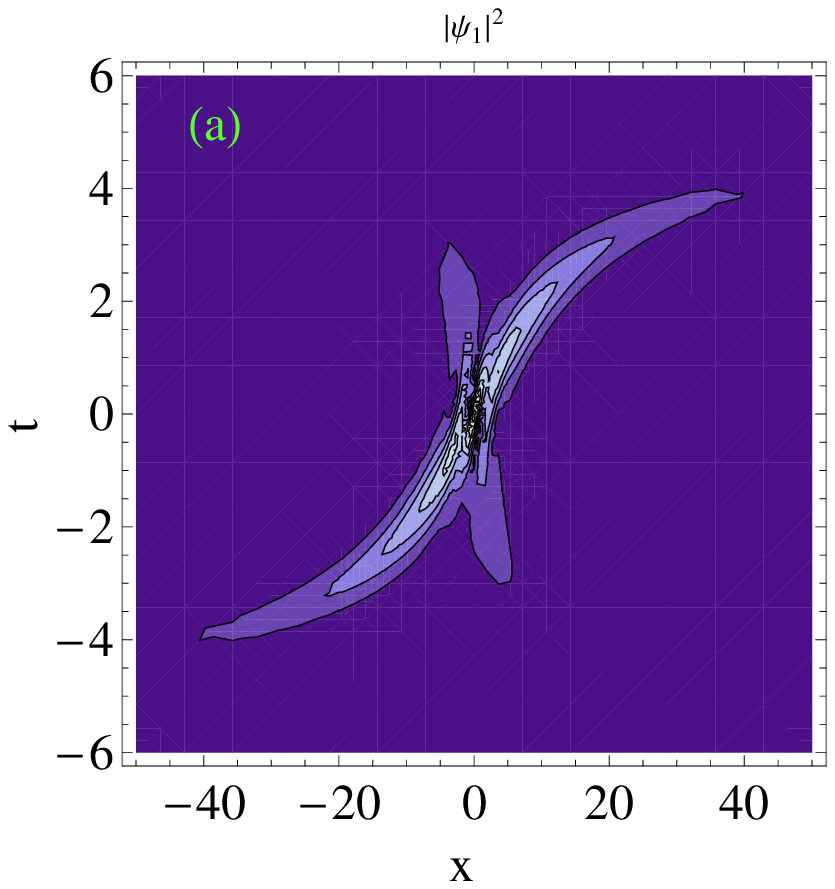,width=0.45\linewidth} \epsfig{file =
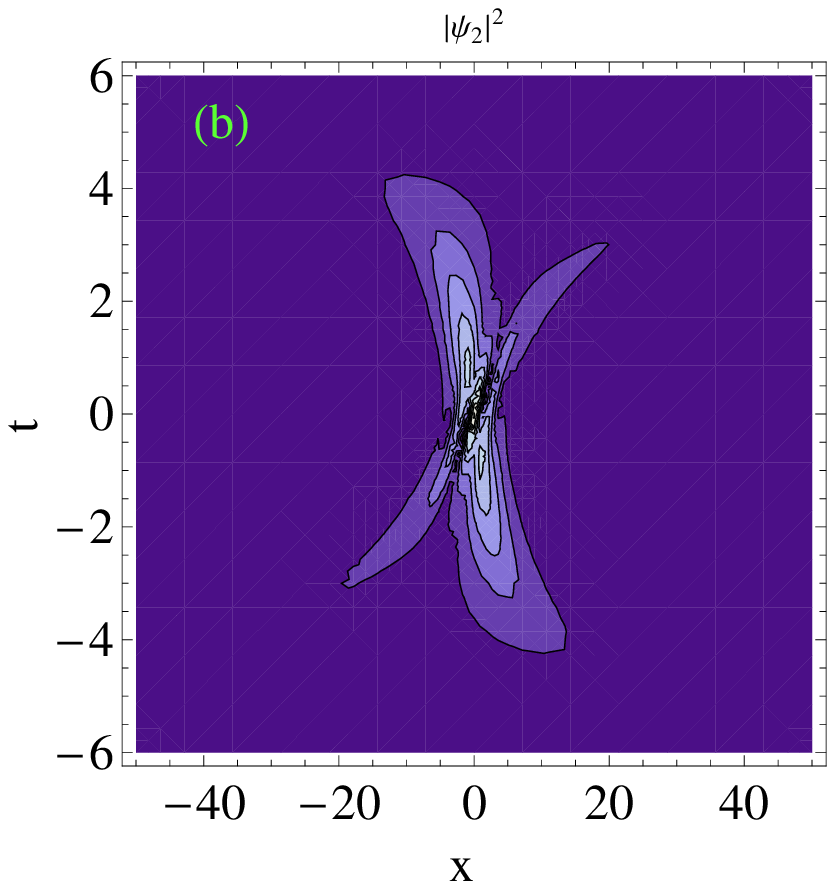, width=0.45\linewidth} \caption{Contour plots of fig
(7)}
\end{figure}
\begin{figure}
\epsfig{file = 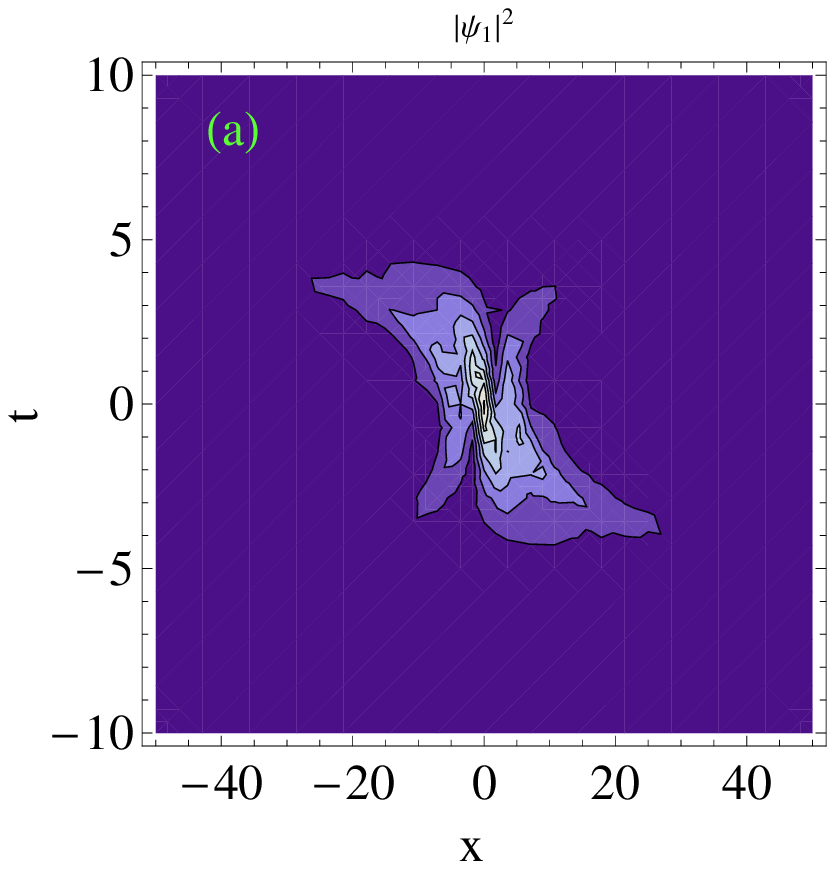,width=0.45\linewidth} \epsfig{file =
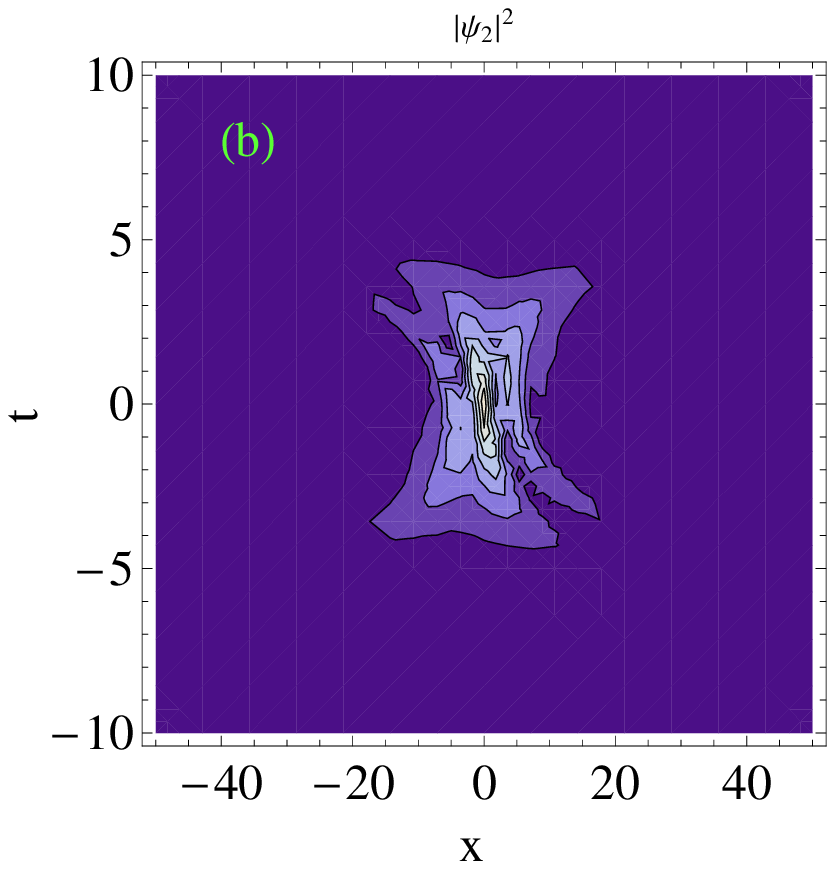, width=0.45\linewidth} \caption{Evolution of two rogue
waves with an increased lifespan by finetuning the trapping
frequency  for $\Gamma(t)=0.1t$}
\end{figure}
\begin{figure}
\epsfig{file = 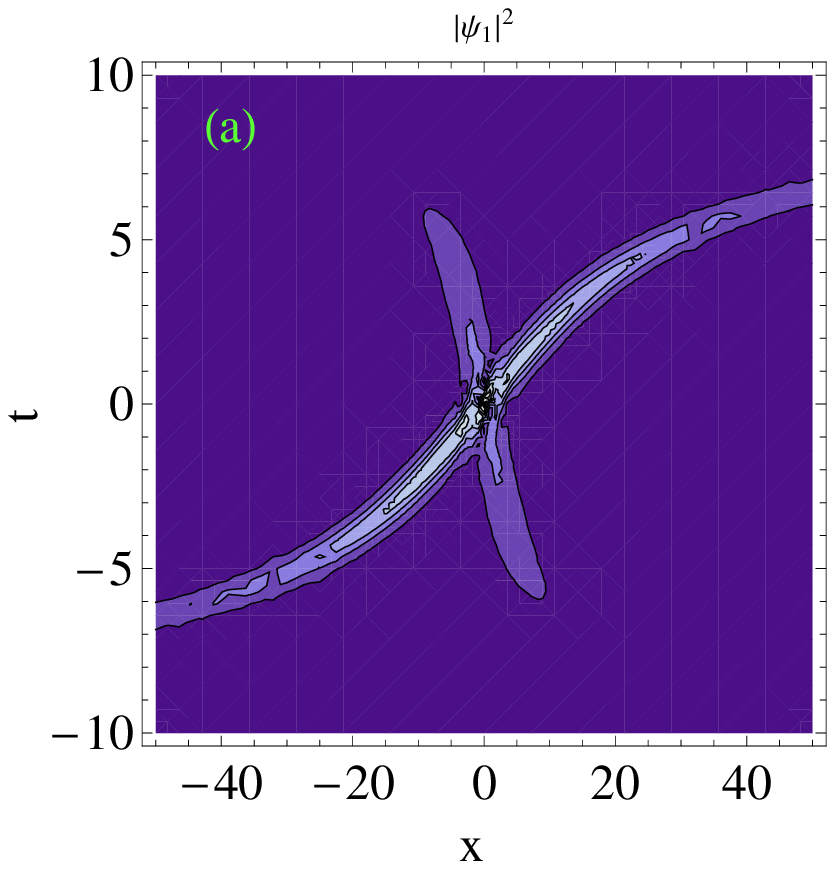,width=0.45\linewidth} \epsfig{file =
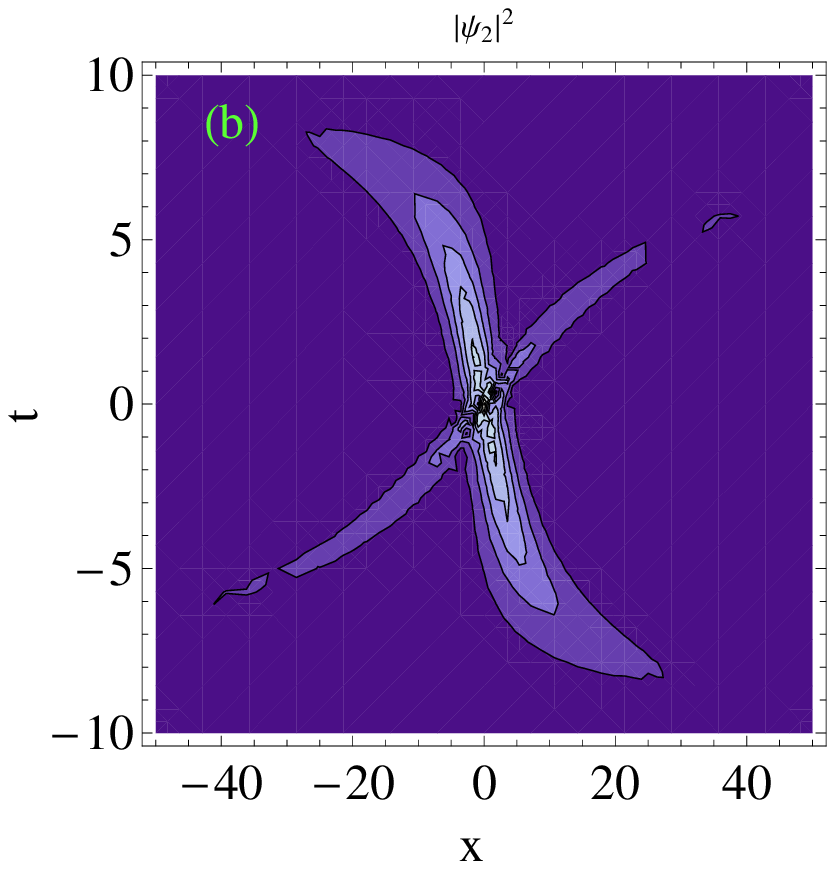, width=0.45\linewidth} \caption{Profile of rogue waves
with an increased lifespan by finetuning the trapping frequency
for $\Gamma(t)=0.03t$}
\end{figure}
\section{Conclusion}
In this paper, we discuss the dynamics of the rogue waves of the
vector BECs governed by the symmetric coupled GP equation. We
observe that we are able to stabilize (or tame) the rogue waves by
either manipulating  the scattering length (both constant and time
dependent) through Feshbach resonance or the trapping frequency.
In the process, we end up increasing the lifespan of rogue waves,
a new phenomenon which may have wider ramifications in BECs and
nonlinear optics.

\textbf{Acknowledgements}:PSV wishes to thank UGC and DAE-NBHM for
the financial support. RR wishes to acknowledge the financial
assistance received from DAE-NBHM (Ref.No:2/ 48(1)/ 2010 / NBHM
/-R and D II/ 4524 dated May.11.2010), UGC (Ref.No:F.No
40-420/2011(SR) dated 4.July.2011) and DST
(Ref.No:SR/S2/HEP-26/2012). KP acknowledges DST and CSIR,
Government of India, for the financial support through major
projects. Authors thank the anonymous referees for their
suggestions to improve the readability of the paper.

\end{document}